\documentclass{icrc29}
\usepackage{graphicx,amssymb,amsmath,times}
\usepackage{epsfig}

\begin{document}

\title[Performance of the  Fluorescence Detector ...]{\bf Performance of the Fluorescence Detectors of the Pierre Auger Observatory}

\author [The Pierre Auger Collaboration]{The Pierre Auger Collaboration \\
         Av. San Mart\'{\i}n Norte 304, (5613) Malarg\"ue, Argentina} 

\presenter{Presenter: J. A. Bellido (jbellido@physics.adelaide.edu.au),\
aus-bellido-J-abs1-he14-oral}

\maketitle 

\begin{abstract}

Fluorescence detectors of the Pierre Auger Observatory
have been operating in a stable manner since January 2004. After
a brief review of the physical characteristics of the detectors,
the associated atmospheric monitoring, the calibration
infrastructure and the detector aperture, we will describe the
steps required for the reconstruction of fluorescence event data,
with emphasis on the shower profile parameters  and primary energy.

\end{abstract}

\section{\bf Introduction}

The Pierre Auger Observatory has been designed with an
order of magnitude increase in collecting power over previous
experiments, coupled with an unprecedented ability to study
detection and reconstruction systematics.  This combination of
statistics and data quality will surely lead to new discoveries
regarding the origin of the highest energy cosmic rays, through
measurements of the energy spectrum, mass composition and arrival
directions.  A unique feature of the Auger observatory is its ``hybrid'' nature,
where showers are observed by surface and fluorescence detectors.

The fluorescence detectors (FD) are distributed in 4 stations around
the perimeter of the surface detector (SD) array, and view the
atmosphere above the array on moon-less or partially moon-lit nights.
Details of the surface and fluorescence detectors are given in
\cite{EApaper}. The full southern Auger array will contain 1600 water
Cherenkov detectors spread over 3000 km$^2$, and currently more than
half of the array is in operation \cite{SDpaper}.  High quality data
are being collected, including ``hybrid'' events described in an
accompanying paper \cite{miguel_paper}.  Here we discuss the
fluorescence detector performance.

\section{\bf The Fluorescence Detectors}

At the present time three of the four fluorescence sites have
been completed and are in operation.  Two of them, Los Leones and
Coihueco, have been collecting data since January 2004, with Los
Morados beginning data taking in March 2005.  The fourth site at
Loma Amarilla will be in operation in the second half of 2006.

A fluorescence site contains six identical fluorescence
telescopes. The telescope design incorporates Schmidt
optics. Fluorescence light enters the telescope through a 1.10 m
radius diaphragm, and light is collected by a 3.5 m $\times$ 3.5 m
spherical mirror and focused onto a photomultiplier (PMT) camera. The
camera contains 440 hexagonal (45 mm diameter) PMTs, each PMT covering
a 1.5$^\circ$ diameter portion of the sky. The optical spot size on
the focal surface has a diameter of approximately 15 mm (equivalent to
0.5$^\circ$) for all directions of incoming light.  To reduce signal
losses when the light spot crosses PMT boundaries, small light
reflectors (``mercedes stars'') are placed between PMTs.  The field of
view of a single telescope covers 30$^\circ$ in azimuth and
28.6$^\circ$ in elevation and an entire fluorescence site (six
telescopes) covers 180$^\circ$ in azimuth and 28.6$^\circ$ in
elevation.  The fluorescence telescopes have been installed
with an uncertainty of 0.1$^\circ$ in their nominal pointing directions.  However, observations of stars
crossing the field of view of the telescopes can improve this
precision, to 0.01$^\circ$.

An optical filter matched to the fluorescence spectrum (approximately
300\,nm to 400\,nm) is placed over the telescope diaphragm to reduce
night-sky noise.  In addition, the diaphragm contains an annular
corrector lens as part of the Schmidt telescope design, with an inner
radius of 0.85\,m and outer radius of 1.10\,m.  The effect of the lens
is to allow an increase in the radius of the telescope diaphragm from
0.85\,m to 1.1\,m (increasing the effective light collecting area by a factor of two) while maintaining an
optical spot size of 0.5$^\circ$ \cite{sato_paper}.

\section{\bf Detector Calibration}
 
One of the goals of the FD is to measure air
shower energies with an uncertainty smaller than 15$\%$. In order to
achieve this goal the fluorescence detectors have to be calibrated
with a precision of about 8$\%$ and the calibration stability needs to
be monitored on a regular basis. An absolute calibration of each
telescope is performed three or four times a year, and relative
calibrations are performed every night during detector operation.

\noindent
{\bf The Absolute Calibration:} To perform an absolute end-to-end
calibration of a telescope, a large homogeneous diffuse light
source was constructed for use at the front of the telescope
diaphragm. This diffuse light source has the shape of a
drum, and has a diameter of 2.5 m. The light flux emitted by the
drum at the diaphragm is known from laboratory measurements
\cite{drum_paper}. The ratio of the drum intensity to the
observed signal for each PMT gives the required calibration. At
present, the precision in the PMT calibration using the drum is
about 12$\%$ \cite{drum_paper}, and is performed at a single
wavelength of 375 nm. The precision will be improved and
calibrations will be performed at other wavelengths.  The drum is
also used to adjust gains for uniform response of the pixels.

\noindent
{\bf Relative Calibration:} Optical fibers bring light signals to
three different diffuser groups for each telescope: (a) at the centre
of the mirror to illuminate the camera; (b) along the lateral edges of
the camera body facing the mirror; (c) facing two reflecting
Tyvek foils glued on the inner side of the telescope shutters. The
total charge per pixel is measured with respect to reference
measurements made at the time of absolute drum calibrations. This
allows the monitoring of short and long term stability, the relative
timing between pixels and the relative gain of each pixel
\cite{aramo_paper}. The relative calibration information is not yet
incorporated in the reconstruction system. However, the average
detector stability has been measured and a corresponding systematic
uncertainty of 3$\%$ has been introduced to account for this.  This
contributes to the overall 12$\%$ systematic uncertainty in the FD
calibration.

\noindent
{\bf Cross-check of the End-to-End Calibration:} Cross-checks of
the FD calibration can be made by reconstructing the energy of
laser beams that are fired into the atmosphere from various
positions in the SD array.  Laser beams are fired to the sky with
known geometry and energy. Part of the laser light is scattered
by the atmosphere (Rayleigh and aerosol scattering) and this
scattered light is detected by the FD telescopes. Using the
measured signal and knowledge of the scattering parameters, it is
possible to estimate the laser energy for comparison with
the real laser energy.  The observed difference between the
reconstructed energy and the real laser energy is of the order of
10$\%$ to 15$\%$ \cite{wiencke_paper}, consistent with the current 
level of uncertainty in calibrations and knowledge of the atmosphere. 

\section{\bf Atmospheric Monitoring}

As part of the reconstruction process, the detected light at the
telescope must be transformed into the amount of fluorescence light
emitted at the shower axis as a function of atmospheric depth.
For this it is necessary to have a good knowledge of local
atmospheric conditions. We need to account for both Rayleigh and
aerosol scattering of light between the shower and the detector,
so we must understand the distribution of aerosols and the
density of the atmosphere at different heights.  
In addition, the temperature distribution with height is needed since the
fluorescence light yield is a (slow) function of temperature.
Finally, the detector volume must be monitored for the presence
of clouds.

\noindent
{\bf Atmospheric Aerosols:} Aerosols in the atmosphere
consist of clouds, smoke, dust and other pollutants. The aerosol
conditions can change rapidly and are known to have a strong
effect on the propagation of fluorescence light. The Observatory
has an extensive network of atmospheric monitoring
devices. These include LIDAR systems, cloud cameras and star
monitors. We have also deployed systems to monitor the wavelength
dependence and differential scattering properties of the
aerosols. More details of these systems can be found in
\cite{roberts_paper}. A steerable laser system, located near the
centre of the Auger array is used to generate tracks that are
seen by the fluorescence detectors \cite{wiencke_paper}. These
tracks also provide a sensitive measure of the aerosol content of
the atmosphere within the aperture of the Observatory. In addition to these monitoring devices, the FD background signal itself is used to measure the aerosol and cloud conditions \cite{caruso}. Presently, only the aerosol information obtained from observing the laser tracks is incorporated in the shower energy reconstruction algorithm.

\noindent
{\bf Molecular Atmosphere:} The atmospheric characteristics at different
heights above the surface array have been studied in 
a number of campaigns with meteorological radiosondings and with continuous
measurements by ground-based weather stations.  In addition,
studies have been made of routine radiosondings performed at
different places in Argentina. Monthly variations are compared to
daily variations as well as year-to-year fluctuations of the
monthly average profiles. The uncertainty in the currently
applied monthly atmospheres in the Auger reconstruction
 introduce an uncertainty in the atmospheric depth at ground
of about 5 g cm$^{-2}$ \cite{bianca_paper}.

\noindent
{\bf Cloud Coverage:} Customized infra-red (7--14 $\mu$m) cloud
cameras have been installed at each fluorescence site. The cloud
camera scans the FD field of view every 5 minutes and the entire sky every 15 minutes, producing sky cloud pictures. The pictures are then processed and 
a database table is filled. The data base indicates which PMTs have their
field of view free of clouds at a given time.

\section{\bf Detector Aperture} 

The response of the Auger fluorescence telescopes has
been simulated in detail and the detector aperture has been
estimated as a function of energy, atmospheric conditions and
primary mass \cite{luis_prado,FDaperture_paper}.  These
fluorescence and hybrid apertures were estimated for a fully
built detector (four fluorescence detectors and a 3000 km$^2$
surface array) and for a detector configuration corresponding to
October 2004 \cite{FDaperture_paper}. For a fully built detector,
the hybrid apertures for cosmic rays with energies of $10^{17.5}$ eV, $10^{18}$ eV,
$10^{18.5}$ eV and greater than $10^{19}$ eV are approximately 900,
3200, 6400 and 7400 km$^2$ sr respectively.

\section{\bf Shower Geometry Reconstruction}

Reconstruction of the shower geometry begins with determination
of the plane containing the shower axis and the FD, the
shower-detector plane or SDP \cite{EApaper}.  The uncertainty in
the reconstructed SDP depends on the size of each PMT field of view ($1.5^\circ$ here),
the observed angular track length, and on the width of the shower
image.  For a typical shower (23$^\circ$ track length) the 
estimated uncertainty (from Monte Carlo) in the normal to the reconstructed 
SDP is about 0.3$^\circ$.

The shower axis within the SDP is defined by the impact parameter
$R_p$, and the angle to the horizontal $\chi_0$.  Reconstruction
of these parameters using the FD alone is prone to difficulty,
especially if the angular extent of the track is small, since a
range of $R_p$ and $\chi_0$ values may fit the measured angular speed
 in the FD camera.  This degeneracy may be broken
with the addition of timing information from a single tank in the
surface array, and the uncertainty of the reconstructed shower
axis is dramatically reduced to approximately 50 m in the core
location and to 0.5$^\circ$ in the shower axis orientation
\cite{carla_paper}.

\section{\bf Shower Profile and Energy Reconstruction}

The signal detected at the FD cameras is converted to the
number of 375\,nm-equivalent photons arriving at the telescope
diaphragm as a function of time \cite{EApaper}. The systematic error
in this transformation is currently 12$\%$ \cite{drum_paper}.

The amount of light emitted at the shower track is calculated
using the shower geometry, the known atmospheric conditions, the
spectrum of the light and the detector's relative wavelength
response.  Using an iterative procedure we take account of the
direct and scattered Cherenkov light measured by the FD \cite{nerling}.  The
resulting fluorescence light at the shower track is converted to
the energy deposited by the shower by applying the expected fluorescence 
efficiency  at each depth.  The overall uncertainty in the transformation from
photons at the detector to emitted fluorescence photons is of the
order of 12$\%$. The application of the fluorescence yield
currently includes a systematic uncertainty on the absolute yield 
of 13$\%$ \cite{nagano_paper}, and  systematics related to the 
pressure (4$\%$), temperature (5$\%$) and humidity (5$\%$).  The 
integral of the shower energy deposit profile provides a calorimetric 
measure of the cosmic ray energy.  A small correction is made for unseen 
energy - an energy-dependent correction of 5-15\% with a systematic
uncertainty of about 3\% \cite{E_correction}.

\section{\bf Conclusions}
Two of the Auger fluorescence detector sites have been
operating in a stable manner since January 2004 and a third site
began operation in March 2005.  Absolute calibration of the FDs
has been performed with a precision of 12$\%$, with improvements
planned to reduce this uncertainty to 8$\%$. The atmospheric conditions 
are constantly monitored and their contribution to the uncertainties 
in the reconstructed shower parameters have been evaluated. The estimated 
systematic uncertainty in the reconstructed shower energy 
is currently 25$\%$, with activity underway to reduce this significantly.

\end{document}